\documentstyle[pra,aps]{revtex}

\begin{document}
\draft

\def\del{\partial}

\title{Tunneling into fractional quantum Hall edges}
\author{Ken-ichiro Imura}
\address{Department of Applied Physics, University of Tokyo,
Tokyo 113-8656, Japan}
\date{\today}
\maketitle

\begin{abstract}
Motivated by the recent experiment by Grayson et.al.,
we investigate a non-ohmic current-voltage characteristics
for the tunneling into fractional quantum Hall liquids.
We give a possible explanation for the experiment
in terms of the chiral Tomonaga-Luttinger liquid theory.
We study the interaction between the charge and neutral modes,
and found that the leading order correction to the exponent 
$\alpha$ $(I\sim V^\alpha)$
is of the order of $\sqrt{\epsilon}$ $(\epsilon=v_n/v_c)$,
which reduces the exponent $\alpha$.
We suggest that it could explain the systematic discrepancy
between the observed exponents and the exact $\alpha =1/\nu$ dependence.
\end{abstract}



Edge tunneling experiments have played a central role
for detecting the non-Fermi liquid properties of fractional quantum
Hall liquids (FQHL).
They include the tunneling between the edges of FQHL
in gated 2D structures \cite{webb},
and the tunneling into the edge of FQHL from a 3D
normal Fermi liquid (FL) \cite{chang}.
Chiral Tomonaga-Luttinger liquid (TLL) \cite{wen1}
predicts non-linear
behaviors of the tunneling conductance for $\nu=1/$(odd integer).
The tunneling conductance through a constricted point contact
scales at low temperatures as $G(T)\sim T^{2/\nu-2}$
when ($eV \ll k_B T$),
whereas the tunneling current into FQHL scales as
$I\sim V^{1/\nu}$ for $eV \gg k_B T$ and
$I\sim T^{1/\nu-1}V$ for $eV \ll k_B T$.
\cite{wen2}
These exponents have been obsreved in the experiments
\cite{webb,chang}, and support the idea that the edge mode of FQH state
is described as a chiral TLL.

For filling factors $\nu=m/(mp+\chi)$ ($m$: integer,
$p$: even integer, $\chi=\pm 1$), \cite{jain}
Kane and Fischer studied the effects of impurity
scattering on the low-energy edge-state dynamics. \cite{kane}
They found a stable fixed point where the phase consists of a single
propagating charge mode and $m-1$ neutral modes.
All the neutral modes propagate at the same speed and manifest
an $SU(m)$ symmetry.
For $\chi=1$ the charge and neutral modes propagate in the
same direction, whereas for $\chi=-1$ they have
different chiralities.
As a result they predicted universal scaling dimensions
for the edge tunneling operators
which correspond to the exponents $\alpha$
for the tunneling into the edge of FQHL ($I\sim V^\alpha$),
\cite{kane}
\[
\alpha = 1/\nu+1-1/m       
=\left\{
\begin{array}{lll}
p+1 & {\rm for} & \chi=1\\
p+1-2/m=2/\nu+1-p& {\rm for} & \chi =-1
\end{array}
\right.
\]
These exponents are consistent with the theory of tunneling
into compressible states, \cite{shytov}
where the non-ohmic $I-V$ characteristics are
explained as a result of the reduction of tunneling density
of states due to the orthogonality catastrophe effect.
However Grayson et. al. studied the power-law behaviors of the $I-V$
characteristics for the tunneling into the edge of FQHL
at various filling factors, and found 
a quasi-linear dependence $\alpha\sim1/\nu$
over a continuum of filling factors $\nu$ from 1/4 to 1.
Their result does not show any strong dependence
on the occurrence or absence of the FQHL. \cite{grayson}
There have been several attempts to resolve the difficulty.
\cite{khvesh,zulicke,dhlee}
In Ref. \cite{khvesh}, it has been proposed that
the effects of unscreened Coulomb interaction
may explain the approximate power-law behavior
reported in Ref. \cite{grayson} for continuously
varing filling factors $\nu$.

One way to explain the experiment by Grayson et.al.
is to work on the chiral TLL. In this case we need a neutral mode
as well as a charge mode to maintain the fermionic commutation relation
of the electrons.
The point is that as far as the electron tunneling is concerned
it is sufficient to focus on only two, a neutral and a charge, edge modes.
\cite{dhlee}
It is true that this ''two-boson model'' cannot be applied to describe 
the quasiparticle physics and that it says nothing about the 
compressibility or incompressibility of
bulk FQHL, but it could be valid independent of the bulk structure
at an arbitrary filling factor $\nu$.
In terms of this two-boson description, the experiment by Grayson et. al.
could be explained in the following way.
Assume that the charge-mode velocity $v_c$ is much larger than the
neutral-mode velocity $v_n$.
This assumption could be justified for the following reason.
The velocities of the edge modes are determined by
the confining potential of our finite $2D$ electron system.
The edge potentials in real samples are usually smooth.
Each edge mode feels an electric field $E$ perpendicular to the
boundary, but its magnitude is proportional to the slope of the confining
potential.
The neutral-mode velocity $v_n$ is roughly given by $E/B$,
which is supposed to be small for a smooth confining potential. 
Whereas for the charge-mode velocity,
the Coulomb interaction plays the role.
If it is screened, $v_c$ is roughly given by
$E/B$ plus the short-range Coulomb interaction,
which could be much larger than $v_n$.
On this assumption one can show that
only the charge mode contributes to the
electron tunneling at the point contact, 
but both charge and neutral modes contribute
to ensure the correct Fermi statistics
of the electron operarors.

In this paper we study the non-linear $I-V$ charcteristics for the
tunneling from a $3D$ FL into a $2D$ FQHL,
in terms of a two-component chiral TLL theory.
We consider both the co-propagating and the counter-propagating edge modes.
In the intermediate step, the structure of the two poles in the
momentum space becomes important,
one corresponding to the charge mode and the other to the neutral mode.
In the absence of interaction and for small $\epsilon=v_n/v_c$,
only the residue at the pole corresponding to the charge mode
contributes to the integration.
In that case we obtain an exponent which is exactly equal to $1/\nu$.
Now we notice the systematic discrepancy between the
observed exponents and the exact $1/\nu$ dependence.
As has been studied by Kane and Fisher, the random equilibration between the
charge and neutral modes tends to make the exponents universal.
Whereas the observed exponents in the experiment by Grayson et. al.
look non-universal, i.e., they are always below the universal line.
Hence we switch off the random impurity scattering, 
and instead we take care of the short-range interaction 
between the charge and neutral modes.
The interaction changes the location of the two poles,
but in a certain region of the parameter space
($\epsilon-\delta$ plane) we have the same global structure of
the two poles as the non-interacting case.
In this paper we mainly work in such a region.
We found that the leading order correction to $\alpha$
is of the order of $\sqrt{\epsilon}$
irrespective of the chiralities of the edge modes.

Our strategy is the following. We start with a $1+1 D$ effective action,
i.e. two-component chiral TLL with a short-range interaction between
the charge and neutral modes. To read effective $K_\rho$ and $K_\sigma$
for the tunneling we integrate out the unimportant degrees of freedom.
For a technical reason, to perform the above procedure we keep both
edge branches for the present,
i.e. both chiralities, but it does not change the physics.
When we discuss the tunneling into FQHL, only the physical
edge contributes.

We begin with the following two-boson model to describe the 
edge mode of FQHL
with a filling factor $\nu\neq 1/$(odd integer),
\[
{\cal L}={\cal L}_c+{\cal L}_n
\]
\begin{eqnarray}
{\cal L}_c &=&
\frac{v_c}{8\pi\nu}
\left[
\left(\frac{\partial\phi_c^+}{\partial x}\right)^2+
\left(\frac{\partial\phi_c^-}{\partial x}\right)^2
\right]
+\frac{i}{4\pi\nu}
\frac{\partial\phi_c^+}{\partial \tau}
\frac{\partial\phi_c^-}{\partial x}
\nonumber \\
{\cal L}_n &=&
\frac{v_n}{8\pi\eta}
\left[
\left(\frac{\partial\phi_n^+}{\partial x}\right)^2+
\left(\frac{\partial\phi_n^-}{\partial x}\right)^2
\right]
+\frac{i}{4\pi\eta}
\frac{\partial\phi_n^+}{\partial \tau}
\frac{\partial\phi_n^-}{\partial x}.
\end{eqnarray}
Here $\phi_{c,n}$ are chiral fields associated with the charge and
neutral modes repectively.
$\phi^\pm=\phi^u\pm\phi^d$ with $\phi^u (\phi_d)$ being the edge mode
propagating in the counter-clockwise (clockwise) direction.
Consider a geometry shown in Fig.~1 of Ref. \cite{grayson}. 
We have a point-like contact between the right edge of FQHL 
(placed on the left side) and the FL (placed on the right side)
through an insulating barrier. 
We assume that the charge mode
propagates in
the counter-clockwise direction, i.e. moves upward in the right edge
of FQHL,
$\phi_c^R=\phi_c^u$.
On the other hand the neutral mode propagates in the clockwise
or counter-clockwise
direction depending on the chirality $\chi$ of the edge modes.
We define $\chi=1$ $(\chi=-1)$ when the neutral mode propagates
upward (downward)
along the right edge of FQHL,
\begin{equation}
\phi_n^R=\left\{
\begin{array}{ll}
\phi_n^u & (\chi=1)\\
\phi_n^d & (\chi=-1)
\end{array}
\right.
\end{equation}
The electron operator at the right edge is given by
\begin{equation}
\Psi_{\uparrow,\downarrow}^R\sim e^{i\phi_c^R/\nu}e^{\pm i\chi\phi_n^R/\eta}
\end{equation}
where $\uparrow,\downarrow$ are quantum numbers corresponding to spin
when $\nu=2/(2p+\chi)$.
The electron Fermi statistics requires
\begin{equation}
{1\over\nu}+{\chi\over\eta}={\rm odd}\ {\rm integer}.
\end{equation}
Only $\phi_R$'s are physical for the tunneling into FQHL,
but it does not change the physics to keep for the time being
the fictitious left-edge modes $\phi_L$.
Now we introduce the short-range interaction $u$ between the charge and
neutral modes.
\begin{equation}
u\left[
\rho_c^L(x)\rho_n^L(x) + \rho_c^R(x)\rho_n^R(x)\right]=
\frac{u}{8\pi^2}\left[
\frac{\partial\phi_c^+}{\partial x}\frac{\partial\phi_n^+}{\partial x}
+\chi
\frac{\partial\phi_c^-}{\partial x}\frac{\partial\phi_n^-}{\partial x}
\right].
\end{equation}
Here the densities are related to the fields as
$\rho=(1/2\pi)\del\phi/\del x$.
Including the interaction between the two edge modes,
the $1+1 D$ effective theory have the following action,
\begin{eqnarray}
S &=& \frac{1}{8\pi\beta l}\sum_{\omega,k}
\phi^T(-\omega,-k)A(\omega,k)\phi(\omega,k)
\nonumber \\
&=& \frac{1}{8\pi\beta l}\sum_{\omega,k}
\left[\phi^{+T},\phi^{-T}\right]
\left[
\begin{array}{ll}
A^+ & B \\
B & A^-
\end{array}
\right]
\left[
\begin{array}{l}
\phi^{+}\\
\phi^{-}
\end{array}
\right],
\end{eqnarray}
where $\phi^T = \left[\phi^{+T},\phi^{-T}\right]
=\left[\phi_c^{+},\phi_n^{+},\phi_c^{-},\phi_n^{-}\right]$,
and $l$ is the length of the edge.
The coefficient matrices are
\[
A^+ (\omega,k)= k^2\left[
\begin{array}{cc}
v_c/\nu    & u/2\pi \\
u/2\pi & v_n/\eta
\end{array}
\right],\ \ \
A^- (\omega,k)= k^2\left[
\begin{array}{cc}
v_c/\nu    & \chi u/2\pi \\
\chi u/2\pi & v_n/\eta
\end{array}
\right]
\]
\begin{equation}
B (\omega,k)= -i\omega k\left[
\begin{array}{cc}
1/\nu & 0 \\
0     & 1/\eta
\end{array}
\right]
\end{equation}
Now we integrate out the unimportant degrees of freedom, i.e. fields except
at the point contact.
In the intermediate step we encounter the following integration, \cite{prb}
\begin{equation}
\int_{-\Lambda_k}^{\Lambda_k}dkA^{-1}(\omega,k)=
\oint dk{\tilde{A}(\omega,k)\over\det A(\omega,k)},
\end{equation}
where $\tilde{A}(\omega,k)$ is the adjugate matrix of
$A(\omega ,k)$.
$\Lambda_k$ is a large-momentum cutoff of the order $1/a$
with $a$ being a lattice constant.
The determinant of $A(\omega,k)$ is calculated to be
\begin{eqnarray}
\det A &=& {k^4\over(\nu\eta)^2}
\left[
\xi^4 k^4-\{(v_c-v_n)^2+2\xi^2\}k^2\omega^2+\omega^4
\right]
\nonumber \\
&=&
{k^4 \xi^4\over (\nu\eta)^2}(k^2+\kappa_c^2)(k^2+\kappa_n^2),
\end{eqnarray}
where
$\xi^2=v_c v_n-\nu\eta(u/2\pi)^2$.

We can choose $\kappa_c$ and $\kappa_n$ such that
$\kappa_n>\kappa_c>0$.
The question is which residues we have to take care of in the
right-hand side of Eq. (8).
In Ref. \cite{dhlee} it is claimed
that when $\Lambda_\omega/v_c\ll\Lambda_k\ll\Lambda_\omega/v_n$
the exponent $\alpha$ is $1/\nu$,
where $\Lambda_\omega$ is a high-frequency cutoff.
This statement might be interpreted
in terms of our contour integration (8)
in the following way.
Let us first switch off the interaction for simplicity.
Then $\kappa$'s reduce to
$\kappa_c=|\omega|/v_c,\kappa_n=|\omega|/v_n$.
If a pole is located far enough from
the path of integration in the left-hand side of Eq. (8),
then the residue at that pole does not contribute to the
contour integration in the right-hand side.
Now we switch on the interaction.
The stability of the system requires that the Hamiltonian should
be positive definite. In our case it reduces to $\xi^2>0$. \cite{prb}
The stability does not allow us to set $v_n$ exactly to zero
for the interacting case. Instead we have two independent variables
to tune the zeros of $\det A$, i.e. the ratio among $v_c, v_n$ and $g$.
We use the following parameterization,
\begin{equation}
{v_n\over v_c}=\epsilon,\ \ \
\nu\eta\left({u\over 2\pi}\right)^2=v_c^2\epsilon(1-\delta),
\end{equation}
where $0\le\delta\le 1$.
When one of the two zeros of $\det A$ in the upper half-plane
goes far away above the physical region, we found that
$\epsilon$ is much smaller than unity for $0\le\delta\le 1$.
Evaluating Eq. (8), we obtain the following effective action,
\cite{prb}
\[
S_\pm[\theta,\lambda]
= \frac{1}{\beta}\sum_{\omega}
\left[
\frac{\pi}{|\omega|}\lambda^{\pm T}(-\omega) P^\pm
\lambda^{\pm}(\omega)
\right.
\left.
+i\lambda^{\pm T}(-\omega)\theta^{\pm}(\omega)
\right],
\]
\begin{equation}
P^+ = \left[
\begin{array}{cc}
p_{c} & -q \\
-q    & p_{n}
\end{array}
\right],\ \ \
P^- = \left[
\begin{array}{cc}
p_{c} & -\chi q \\
-\chi q    & p_{n}
\end{array}
\right]
\end{equation}
where $\lambda$'s are Lagrange multipliers introduced
for integrating out the $\phi$'s.
Leading order terms in the expansion w.r.t. $\epsilon$ are
obtained as
\begin{eqnarray}
p_c &=& \nu [1+(1-\delta)\epsilon+{\cal O}(\epsilon^2)], 
\nonumber \\
p_n &=& \eta(1-\delta)
[\epsilon+(1-3\delta)\epsilon^2+{\cal O}(\epsilon^3)],
\nonumber \\
q &=& \sqrt{\nu\eta(1-\delta)\epsilon}
[1+(1-2\delta)\epsilon+{\cal O}(\epsilon^2)].
\end{eqnarray}
We still have an interaction between the charge and neutral modes,
but $\lambda_+$ and $\lambda_-$ are decoupled.
One might notice that $\det P$ vanishes.
Hence it would be better to work in the base, 
\[
\left[
\begin{array}{c}
\theta^\pm_1\\
\theta^\pm_2
\end{array}
\right]
=L^\pm
\left[
\begin{array}{c}
\theta^\pm_c\\
\theta^\pm_n
\end{array}
\right],
\]
\begin{equation}
L^+={1\over\sqrt{p_c^2+q^2}}\left[
\begin{array}{cc}
p_c & -q   \\
q  & p_c
\end{array}
\right],\ \ \ 
L^-={1\over\sqrt{p_c^2+q^2}}\left[
\begin{array}{cc}
p_c & -\chi q   \\
\chi q  & p_c
\end{array}
\right]
\end{equation}
where $P_\pm$ is diagonalized, 
$L^{\pm} P^\pm L^{\pm T}={\rm diag}[p_c+p_n,0]$
Integration over $\lambda_2^\pm$ just gives
a constraint on $\theta_2^\pm$.

After integrating out all the unimportant degrees of freedom
we obtain a $0+1 D$ effective theory. Together with the cosine
term which describes the tunneling, the phase $\theta$
could be regarded as a coordinate of a quantum Brownian
particle moving in the periodic potential and coupled to
a dissipative environment. \cite{caldeira}
Now the power law behavior of the electron Green's function
at the point contact is controlled by the strength of dissipation.
Hence the coefficient of the dissipation term is proportional
to the exponent $\alpha$.
For a strong (weak) dissipation
the $I-V$ characteristics tends to show a  sub-ohmic (super-ohmic)
behavior.
To discuss the Grayson's experiment we calculate
the corrections to $K_\rho$ and $K_\sigma$
up to the first order in $\epsilon$.
The dissipation term reads
\begin{equation}
S_{diss}[\theta] =
{1\over 4\pi\beta}\sum_{\omega}{1\over p_c+p_n}|\omega|
\left(
|\theta_1^+(\omega)|^2 +
|\theta_1^-(\omega)|^2
\right)
\end{equation}
The electron Green's function at the point contact
decays as
\begin{equation}
G_{\uparrow, \downarrow}(\tau) =\langle
\Psi_{\uparrow, \downarrow}^R(\tau)
\Psi_{\uparrow, \downarrow}^{R\dagger} (0)\rangle
\sim
e^{-\left[g_{c}(\tau)/\nu^2+g_{n}(\tau)/\eta^2
\pm\chi 2f(\tau)/\nu\eta \right]},
\end{equation}
where $g_c=g_{cc},g_n=g_{nn}$ and $f=g_{cn}$ are
the boson correlation functions,
\begin{equation}
g_{ij}(\tau) = \langle\theta_{i}^{R}(\tau)\theta_{j}^{R}(0)\rangle
\sim Q_{ij}\ln {1\over\tau},
\end{equation}
where $i,j=c,n$.
One finds
\begin{equation}
Q = \left[
\begin{array}{cc}
p_{c} & -q \\
-q    & p_{n}
\end{array}
\right]
\end{equation}
for both chiralities.
The electron operator which has the lower exponent is dominant
and determines the exponent $\alpha$.
Up to the 1st order in $\epsilon$ we obtain
\begin{equation}
\alpha=
{1\over\nu}-2\sqrt{{(1-\delta)\epsilon\over\nu\eta}}+
\left({1\over\nu}+{1\over\eta}\right)(1-\delta)\epsilon
\end{equation}
We found a correction of the order of $\sqrt{\epsilon}$
to the exponent $\alpha$ for both chiralities.
The leading order correction comes from
the residual interaction between the charge and neutral modes.
Eq. (18) reproduces the result in Ref. \cite{dhlee}
in the limit $\delta=1$ where the interaction vanishes.

The above result is quite different from the case where both of 
the poles contribute to the tunneling. In that case,
the exponent is robust against the interaction for $\chi=1$,
\begin{equation}
\alpha={1\over\nu}+{1\over\eta},
\end{equation}
whereas for $\chi=-1$,
\cite{prb}
\begin{equation}
\alpha={2K_\rho^{\rm eff} \over \nu^2}+{K_\sigma^{\rm eff}\over \eta^2}
-{2g\over\nu\eta},
\end{equation}
where
\[
K_\rho^{\rm eff} = {\nu/2\over
\sqrt{1-{\nu\eta\over\pi^2}\left({u\over v_c+v_n}\right)^2}},\ \ \
K_\sigma^{\rm eff} = {1\over
\sqrt{1-{\nu\eta\over\pi^2}\left({u\over v_c+v_n}\right)^2}}.
\]
\begin{equation}
g=\sqrt{\nu\eta\over {\pi^2\over \nu\eta}\left({v_c+v_n\over u}\right)^2-1}.
\end{equation}

Let us compare our results with the experiment.
One might read the exponent $\alpha$ from the experiments as 
\cite{chang,grayson}
\begin{equation}
\alpha\sim\left\{
\begin{array}{lllll}
1.3 & {\rm for} & \nu=2/3, & \eta=2 & (\chi=-1)\\
2.3 & {\rm for} & \nu=2/5, & \eta=2 & (\chi=1)
\end{array}
\right..
\end{equation}
We obtain
$(1-\delta)\epsilon\sim 4\times10^{-3}$
for $\nu=2/3,\eta=2$, and $(1-\delta)\epsilon\sim 2\times10^{-3}$
for $\nu=2/5,\eta=2$, 
which are consistent with the assumption that $\epsilon\ll 1$
if the interaction is not too small.
Let us mention that the observed exponents
are always smaller than $1/\nu$. \cite{grayson}
We suggest that the short-range interaction
between the charge and spin modes could
explain the discrepancy between the observed
exponents in the experiment and the exact
$\alpha=1/\nu$ dependnece for continuously
varying filling factors.

Finally we comment on the effect of unscreened Coulomb
interaction. Consider a sample with width $w$ which is much
larger than the distance $d$ between the 2D FQHL and 3D FL.
Assume that $k_B T \ll eV$, i.e. the high-energy cutoff of the
system be $\Lambda_\omega \sim eV$.
As the voltage is decreased the unscreened Coulomb interaction
becomes important when $eV \ll 1/d$.
For the voltages $1/w \ll eV \ll 1/d$ the long-range Coulomb
interaction which comes from the FQH edge running paralell to
the edge of the 3D region is important.
This interaction repels the electron from the point contact,
hence decreasing $K_\rho$. As a result the long-range Coulomb
interaction enhances the tunneling current in this energy scale.
However as the voltage is further decreased the long range
Coulomb interaction extended to the left side (far from
the barrier) becomes
dominant when $eV \ll 1/w$.
This interaction repels the electron from the extended part of
the edges and effectively attracts the electron to a point contact,
hence increasing $K_\rho$.
The long-range Coulomb interaction suppresses the tunneling current
when $eV \ll 1/w$, and in the end the conductance decays 
exponentially when both the temperature and the voltage goes to
zero. \cite{ssc}

In conclusion
we investigated the effects of interaction
on the tunneling into FQHL at continuously varying filling factors.
We found that the leading order correction is of the
order of $\sqrt{\epsilon}$
which reduces the exponent $\alpha$.
We suggest that it may resolve a discrepancy between the
observed exponents and the exact $\alpha =1/\nu$ dependence.

\acknowledgements
The author is grateful to Albert Chang, Brad Marston,
K.-V. Pham and Pascal Lederer
for useful discussions. He is also grateful
to Naoto Nagaosa for his suggestions and encouragement.
This work was supported by COE and Priority Areas Grants
from the Ministry of Education, Science and Culture of
Japan. He is also supported by JSPS Research
Fellowships for Young Scientists.

\end{document}